\documentclass[aps,pre,twocolumn,superscriptaddress,showpacs,floatfix]{revtex4-1}
\usepackage{graphicx}
\usepackage{amssymb,amsmath,amsbsy,amsfonts,bm}
\usepackage{epsfig}
\usepackage{subfigure}
\usepackage{multirow}
\usepackage{threeparttable}
\usepackage[percent]{overpic}
\usepackage{color}
\definecolor{darklavender}{rgb}{0.45, 0.31, 0.59}
\definecolor{amethyst}{rgb}{0.6, 0.4, 0.8}

\begin{document}
\title{ {Active Brownian and inertial particles in disordered
 environments: \\
short-time expansion of the mean-square displacement}}
\author{Davide Breoni}
\affiliation{Institut f\"ur Theoretische Physik II: Weiche Materie, Heinrich
Heine-Universit\"at D\"usseldorf, Universit\"atsstra{\ss}e 1, 
40225 D\"usseldorf, Germany}
\author{Michael Schmiedeberg}
\affiliation{Institut f\"ur Theoretische Physik 1, Friedrich-Alexander-Universit\"at Erlangen-N\"urnberg, Staudtstra{\ss}e 7, 
91058 Erlangen,
 Germany}
\author{Hartmut L\"owen}
\affiliation{Institut f\"ur Theoretische Physik II: Weiche Materie, Heinrich
Heine-Universit\"at D\"usseldorf, Universit\"atsstra{\ss}e 1, 
40225 D\"usseldorf, Germany}
\begin{abstract}
We consider an active Brownian particle moving in a disordered two-dimensional
 energy or motility landscape.
The averaged mean-square-displacement (MSD) of the particle is calculated analytically within
 a systematic short-time expansion. As a result, for overdamped particles, 
 both
 an external random force field and  disorder in the self-propulsion speed
 induce ballistic behaviour adding to the  ballistic 
regime of an active particle with sharp self-propulsion speed. Spatial correlations in the 
force and motility landscape contribute only to the cubic and higher order
powers in time for the MSD. Finally, for inertial particles
two superballistic regimes are found where the scaling exponent of the MSD with time is $\alpha=3$ and  $\alpha=4$.
 We confirm our theoretical predictions by computer simulations. Moreover they  are verifiable in experiments on self-propelled colloids in random environments.

\end{abstract} 
\maketitle

\section{Introduction}
\label{Sec:Introduction}
The motion of active colloidal particles in complex environments is a
vivid topic of recent physics research \cite{bechinger_active_2016,reichhardt_depinning_2017,gompper_2020_2020}. In particular if self-propelled particles are moving
in a heterogeneous or random medium, there is a plethora of new effects created by disorder.
Examples include trapping and clogging of particles \cite{chepizhko_diffusion_2013,reichhardt_clogging_2018,reichhardt_avalanche_2018}, destruction of flocks \cite{morin_distortion_2017}, the
control of crowds \cite{pince_disorder-mediated_2016,koyama_separation_2020} and subdiffusive long-time dynamics \cite{chepizhko_diffusion_2013,bertrand_optimized_2018,dor_ramifications_2019,morin_diffusion_2017}.
The random environment can be established by a porous medium \cite{grancic_active_2011,blagodatskaya_active_2013},
by fixed obstacle particles \cite{takagi_hydrodynamic_2014,lozano_active_2019,jin_fine_2019,mokhtari_collective_2017,alonso-matilla_transport_2019,brun-cosme-bruny_deflection_2020}
or by optical fields (such as a speckle field \cite{volpe_brownian_2014,volpe_speckle_2014,bewerunge_experimental_2016,nunes_ordering_2020,pesce_step-by-step_2015,paoluzzi_run-and-tumble_2014,bianchi_active_2016})
which can create both random external potentials  \cite{bewerunge_colloids_2016,bewerunge_time-_2016,hanes_brownian_2013,evers_colloids_2013,hanes_colloids_2012,stoop_clogging_2018,chaki_escape_2020} or
a motility landscape \cite{lozano_phototaxis_2016,lozano_propagating_2019}.

While the control of particle motion in a random environment is crucial for many applications
such as steered drug delivery and minimal invasive surgery,
also the fundamental physics needs to be understood within statistical mechanics. In particular,
analytical solutions for simple model systems are important here to unravel the underlying principles.
A particular successful model for self-propelled particles is that of active Brownian motion
\cite{howse_self-motile_2007,ten_hagen_brownian_2011,lowen_inertial_2020} designed for colloidal microswimmers. Basically the particle
 performs overdamped  motion under the action of an internal effective drive  directed
 along its orientation which is experiencing Brownian fluctuations establishing a persistent
random walk of the particle. In this model,
the MSD of the particle exhibits a crossover from ballistic
behavior governed by directed self-propulsion to final long-time diffusion with a diffusion coefficient that scales
with the square of the self-propulsion velocity.
The motion of self-propelled particles in various random environments has been studied  by using computer simulations
of active Brownian particles or related models
 \cite{chepizhko_diffusion_2013,chepizhko_active_2015,chepizhko_optimal_2013,
 schirmacher_anomalous_2015,chepizhko_ideal_2019,chepizhko_random_2020,
 reichhardt_active_2014,reichhardt_aspects_2014,kumar_symmetry_2011,
 kumar_flocking_2014,das_polar_2018,quint_swarming_2013,simon_brownian_2016,
 zhu_transport_2018,ai_flow_2019,sandor_dynamic_2017,zeitz_active_2017,jakuszeit_diffusion_2019}.
Also some experiments for active particle in disordered landscapes
 have been performed on colloids \cite{volpe_microswimmers_2011,morin_distortion_2017,pince_disorder-mediated_2016,lozano_active_2019} and bacteria \cite{bhattacharjee_confinement_2019}.
However, analytical results are sparse, even for a single active particle. In one spatial
dimension, exact results have been obtained for a run-and-tumble particle \cite{dor_ramifications_2019}. In higher dimensions, analytical results are available for
discrete lattice models \cite{bertrand_optimized_2018} and for a highly entangled slender self-propelled rod  \cite{mandal_crowding-enhanced_2020,romanczuk_active_2012}.

Here we present analytical results for the off-lattice model of active Brownian motion in two dimensions
by exploring the short-time behavior of the mean-square-displacement. The self-propelled particle is experiencing
a space-dependent landscape of quenched disorder \cite{bouchaud_anomalous_1990,duan_breakdown_2020} of an external force or the internal motility field.  We calculate
the averaged mean-square-displacement (MSD) of the particle  for arbitrary disorder strength
in a systematic short-time expansion. 
 As a result, for overdamped particles, randomness in the external force field and the particle motility
both contribute to the initial ballistic regime. Spatial correlations in the 
force and motility landscape contribute only to the cubic and higher order
powers in time for the MSD. Finally, for inertial particles
which are initially almost at rest  three subsequent regimes can occur where the scaling exponent of the MSD with time crosses over from an initial $\alpha=2$ to a transient $\alpha=3$ and a final  $\alpha=4$. 
The latter superballistic regimes are traced back to the initial acceleration.
We remark that similar superballistic exponents have been found for an active Brownian
particle in linear shear flow \cite{ten_hagen_brownian_2011} and for animal motion \cite{tilles_random_2017}
but the physical origin is different in these cases.
Our  predictions are confirmed by computer simulations and are in principle verifiable in experiments on self-propelled colloids
in random environments.

As an aside, we also present results for a passive particle in an random force landscape. Note that we consider the short-time behavior that is also briefly mentioned in \cite{bewerunge_colloids_2016,bewerunge_time-_2016,hanes_brownian_2013,evers_colloids_2013,hanes_colloids_2012,wilkinson_flooding_2020,Zunke_PhD_Thesis} though in these works usually the focus is on the long-time behavior  \cite{bewerunge_colloids_2016,bewerunge_time-_2016,hanes_brownian_2013,evers_colloids_2013,hanes_colloids_2012,Zunke_PhD_Thesis} or the mean first passage time \cite{wilkinson_flooding_2020} of such systems.

The paper is organized as follows: in the next section
we discuss the model of a single Brownian particle interacting with an external random landscape, in the subsequent one we move on to the case of a random motility field and in both cases we consider both an overdamped and an underdamped particle.
Finally in Sec. \ref{Sec:conclusions} we conclude with a summary of our results and possible continuations of our work.

\section{Active particle in a disordered potential energy landscape}
\label{Sec:potential}

\subsection{Overdamped active Brownian motion}

\label{Sub:OPot}
We start by considering a single active Brownian particle moving in the two-dimensional plane. The dynamics is 
assumed to be overdamped as relevant for micron-sized swimmers and self-propelled colloids at low Reynolds number. 
The position of the particle center is described by its
trajectory ${\vec{r}}(t)=(x(t),y(t))$ and its orientation is given by a unit vector $\hat{u}(t)=(\cos \phi (t), \sin \phi (t) )$
where $\phi$ is the angle of the orientation vector with the $x$-axis and $t$ is the time. The equations of motion of an overdamped  active Brownian particle
for the translation and rotation degrees of freedom are given by

\begin{align}
\label{1}
\gamma\dot{\vec{r}}(t)&= \gamma v_0\hat{u}(t)+\vec{f}(t)+\vec{F}(\vec{r}(t)),\\
\label{2}
\gamma_R\dot{\phi}(t)&= f_R(t),
\end{align}
where $\gamma$ and $\gamma_R$ are, respectively,
 the translational and rotational friction coefficients and $v_0$ is the self-propulsion velocity 
which is directed along the orientation vector $\hat{u}(t)$. The terms 
$\vec{f}(t)$ and $f_R(t)$ represent Gaussian white noise forces and torques originating from the solvent kicks with
\begin{align}
\label{3} \langle \vec{f}(t) \rangle &= 0, \\
\label{4} \langle f_i(t)f_j(t') \rangle &= 2D \gamma^2\delta(t-t')\delta_{ij},\\
\label{5} \langle f_R(t) \rangle &= 0, \\
\label{6} \langle f_R(t)f_R(t') \rangle &= 2D_R \gamma_R^2\delta(t-t').
\end{align}
Here $\langle \cdot \rangle$ is the thermal noise average, $D$ is the translational free diffusion constant and $D_R$ is the rotational one.\\
Importantly, the particle is exposed starting at $t=0$ to an external force field $\vec{F}(\vec{r})$ representing the static quenched disorder. 
We assume that the external force is conservative, i.e.\ that it can be derived as a gradient from a random potential energy $V(\vec{r})$ such that 
\begin{equation}
\label{9}
\vec{F}(\vec{r})=-\vec{\nabla}V(\vec{r})
\end{equation}
holds. For the scalar potential energy we choose a general decomposition into two-dimensional Fourier modes and assume that the amplitudes in front of these modes are randomly Gaussian distributed and uncorrelated.
In detail, the random potential $V(\vec{r})$ is expanded as
\begin{align}
\label{10}V(\vec{r})=-\sum_{i,j=0}^\infty\left(\epsilon_{ij}^{(1)}\cos(k_ix+k_jy)+\epsilon_{ij}^{(2)}\sin(k_ix+k_jy)\right),
\end{align}
where $k_n=\frac{2\pi}{L}n$, $L$ denoting a large periodicity length. 
The amplitudes $\epsilon_{ij}^{(\alpha)}$ are Gaussian random numbers which fulfil
\begin{align}
\label{11}\overline{\epsilon_{ij}^{(\alpha)}}=0 \text{~~~and~~~} \overline{\epsilon_{ij}^{(\alpha)}\epsilon_{mn}^{(\beta)}}=\overline{\epsilon_{ij}^{(\alpha)2}}\delta_{im}\delta_{jn}\delta^{\alpha \beta},
\end{align}
where $\overline{(\cdot)}$ denotes the disorder average. We further assume the potential to be isotropic, meaning that the $\epsilon_{i,j}$ are only functions of $i^2+j^2$.\\
Now we compute the mean-square-displacement (MSD) $\Delta(t)$ of the particle which is initially at time $t=0$ at position ${\vec r}_0$ 
with orientational angle $\phi_0$. In this paper, we consider a disorder-averaged MSD, in detail
it is a {\it triple\/} average over i) the thermal noise $\langle \cdot \rangle$, ii) the disorder $\overline{(\cdot)}$, and iii) the initial conditions $\ll \cdot \gg$. As we switch on the potential at $t=0$, due to translational invariance and self-propulsion isotropy, the latter are assumed to be homogeneously distributed
 in space and in the orientational angle. 
Consequently,
\begin{align}
\label{12}
    \Delta(t)& := \ll \langle\overline{(\vec{r}(t)-\vec{r}_0)^2}\rangle \gg.
\end{align}
In order to simplify the notation, the average over both disorder and initial conditions for the various components and derivatives of the forces will be abbreviated by the symbol $\widehat{(\cdot)}$, for example $\ll \overline{F^2_x(\vec{r}_0)} \gg\equiv \widehat{F^2_x}$.

In Appendix A, we detail the analytical systematic short-time expansion in terms of powers of time $t$ for the MSD.
Up to fourth order, the final result reads as
\begin{align}
\label{13}
    \Delta(t)&=4Dt+\left[v_0^2+\frac{1}{\gamma^2}\widehat{F_i^2}\right]t^2-\left[\frac{1}{3}v_0^2D_R+\frac{D}{\gamma^2}\widehat{F_i^{j2}}\right]t^3\nonumber\\
    &+\frac{1}{24}\left[2v_0^2D^2_R+10 \frac{D^2}{\gamma^2}\widehat{F_i^{jk2}}-5\frac{v_0^2}{\gamma^2}\widehat{F_i^{j2}} \right.\nonumber\\
    &+\frac{1}{\gamma^4}\left(14\widehat{F_i^2F_i^{i2}}+8\widehat{F_i^3F_i^{ii}}+14\widehat{F_xF_yF_x^{y}F_i^{i}}\right.\nonumber\\
    &\left. \left. +14\widehat{F_yF_xF_y^{x}F_i^{i}} -5\widehat{F_i^2F_x^{y2}}-5\widehat{F_i^2F_y^{x2}}\right)\right]t^4+\mathcal{O}\left( t^5\right).
\end{align}
Here our convention in the notation is that the presence of any index $i$, $j$ or $k$ implies an additional sum over the directions $x$ and $y$.
For example, in this compact notation, we have $\widehat{F_i^2}\equiv \sum_{i=x,y}\widehat{F_i^2}$.  Subscripts in $F$ indicate the Cartesian component of the force, while superscripts denote a spatial derivative. For example, $\widehat{F_i^{j2}}= \sum_{i=x,y} \sum_{j=x,y} \widehat{ (\frac {\partial F_i}{\partial j})^2 }$.\\
In order to assess the presence of scaling regimes for the MSD, it is necessary to know if the prefactors of $t^\alpha$ are negative or positive, and hence what is the sign of the various force products. In Eq.(\ref{13}), it can be shown that all products are positive with the exception of $\widehat{F_i^3F_i^{ii}}$. In the special case of a single mode potential, that we define as a potential where only $\epsilon_{11}\neq 0$, one can simplify this negative product with all the ones with $1/\gamma^4$ prefactor and obtain the shorter and positive expression $6\widehat{F_i^2F_j^{k2}}$ (see Appendix \ref{Appendix}). In the more general case positivity is not ensured.\\
Let us now discuss the basic result contained in Eq.(\ref{13}). First of all, in the absence of any external forces, we recover the analytical expression  for a
free active particle \cite{howse_self-motile_2007} where \begin{align}
\label{14}
    \Delta(t)&=4Dt+2\frac{v_0^2}{D_R^2}\left( D_Rt+\text{e}^{-D_Rt}-1\right) \nonumber\\
    &=4Dt+v_0^2t^2-\frac{1}{3}v_0^2D_Rt^3+\frac{1}{12}v_0^2D^2_Rt^4+\mathcal{O}\left( t^5\right)
\end{align}
expanded up to order $\mathcal{O}\left( t^5\right)$.
Conversely, for finite forces but in the limit of no activity, $v_0=0$, we get results for a passive 
particle in a random potential energy landscape \cite{Zunke_PhD_Thesis}.

In general, for both $v_0 \not= 0$ and ${\vec F} \not= 0$, as far as the influence of disorder is concerned, 
the first leading correction in the MSD is in the ballistic $t^2$-term. The physical interpretation of this term is rooted in the fact that in a disordered energy landscape on average the particle actually feels a non-vanishing force such that it is drifting. The resulting
ballistic contribution is on top of the activity itself which  also contributes to the transient ballistic regime. We define now the crossover time $t^c_{1\rightarrow 2}$ as the ratio $A_1/A_2$ between the two regimes scaling with $A_1t$ and $A_2t^2$. This quantity indicates the time when the ballistic regime becomes prominent over the diffusive one. In this case $t^c_{1\rightarrow 2}$ depends on the self-propulsion velocity and the strength of the potential, and more specifically it shrinks as those grow:
\begin{equation}
\label{15}
t^c_{1\rightarrow 2}=\frac{4 D}{\widehat{F^2_i}/\gamma^2+v_0^2},
\end{equation}
meaning that an active particle subject to a random force field begins earlier to move ballistically. Spatial correlations in the random potential energy landscape are contributing to the $t^3$-term in lowest order 
and affect the higher powers in time as well. 
Clearly, from the result  (\ref{13}), the prefactor in front of the $t^3$-term is negative such that there is no regime where a pure $t^3$-scaling in the MSD can be observed. 

Finally, one could deduce from Eq.(\ref{13}) that there is a special limit of parameters where the dominant regime is an acceleration  where $\Delta(t) \propto t^4$. In order to see this, one can set $v_0$ and $D$ to be small, while considering large wave vectors $k$ and amplitudes $\epsilon$ in the potential decomposition Eq.(\ref{10}) such that any combination of $\epsilon^2 k^4$ is much larger than one. However, this is not a scaling regime, as the term $\mathcal{O}\left(t^6\right)$ dominates on $\mathcal{O}\left(t^4\right)$ in the same limit.

\begin{figure}[!htbp]
 \begin{center}
 \includegraphics[width=8.7cm]{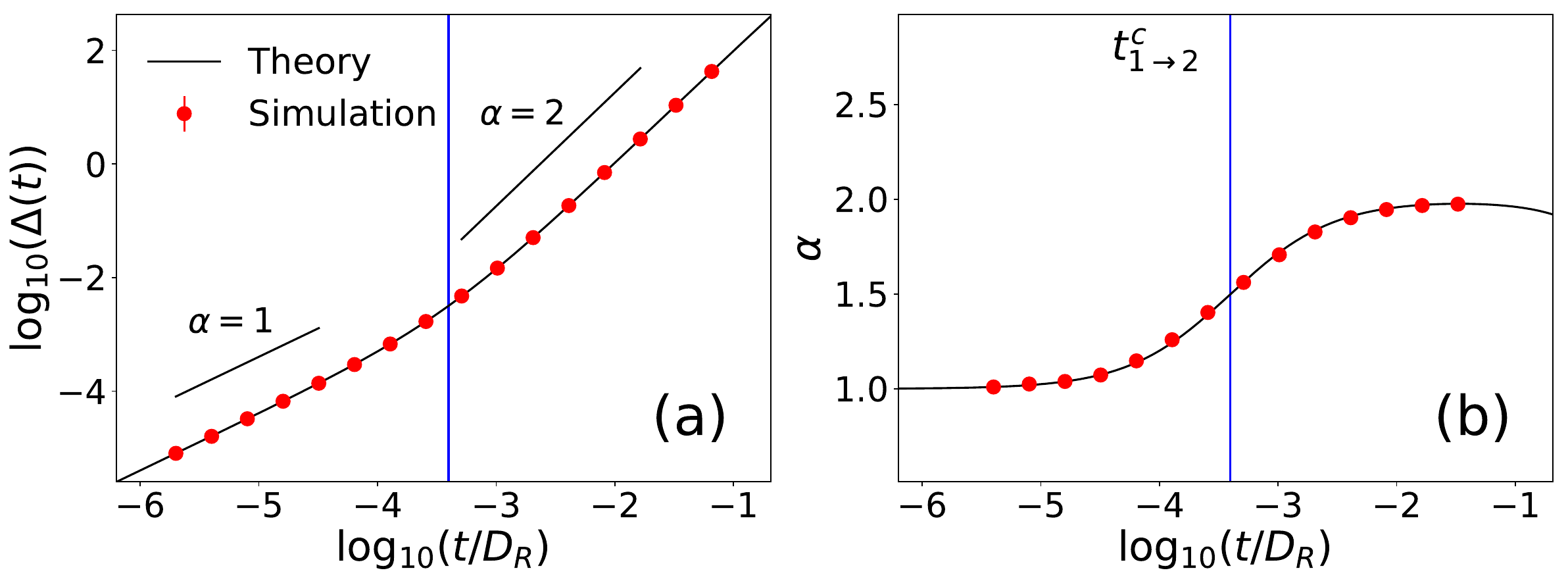}
 \includegraphics[width=8.7cm]{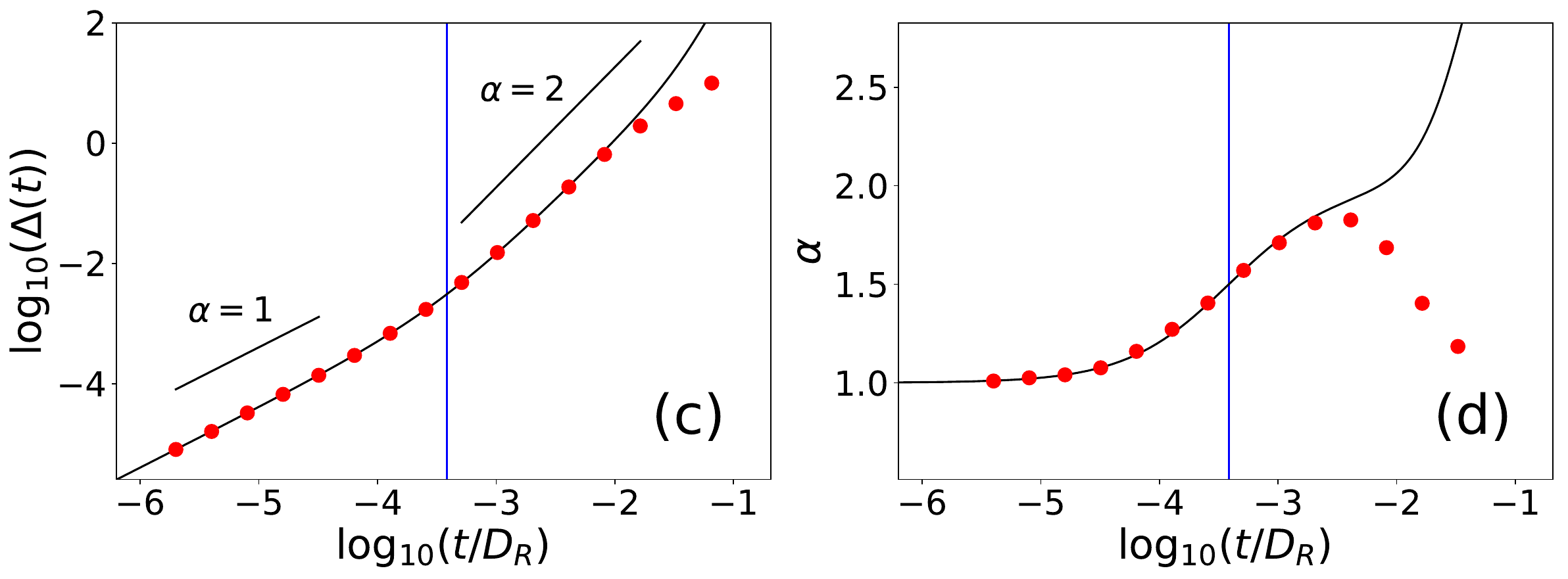}
 \caption{Mean square displacement (a,c), scaling exponent $\alpha$ (b,d) and crossing time $t^c_{1\rightarrow 2}$ (marked by a blue line) for an overdamped active particle in a random single mode potential. In (a,b) we used the parameters $v_0=100\sqrt{DD_R}$, $\epsilon=100D\gamma$ and $L=100\sqrt{D/D_R}$. As described by the theory, the initial diffusive behavior is soon replaced by the ballistic behavior. In (c,d) the parameters $v_0=50\sqrt{DD_R}$, $\epsilon=100D\gamma$ and $L=10\sqrt{D/D_R}$ also show first the diffusive and then the ballistic regimes, but for larger times the short-time expansion approximation breaks down earlier, as the average $\epsilon^2 k^4$ is larger.}
\label{Fig1}
\end{center}
\end{figure}

We compared the result (\ref{13}) to standard Brownian dynamics computer simulations. 
In our simulations, we first generated a random energy landscape, then the particle was exposed to the selected landscape with an initial random position and orientation. Then we integrated the equations of motion with a Euler finite difference scheme involving a finite time step of typically $\Delta t = 10^{-6}/D_R$. In order to simplify calculations for the simulations, we always used single mode potentials.
The MSD was then appropriately averaged over many starting configurations, the number of which was always larger than $10^4$. This amount was large enough to yield statistical errors always below 1\% of the averaged values of the MSD. We believe these samples are hence large enough to ensure ergodicity for the initial conditions.\\ 
Figure \ref{Fig1} shows  examples for the scaling behavior of both the MSD and its scaling exponent 
\begin{equation}
\label{16}
\alpha(t) := \frac{d(\log(\Delta(t)))}{d(\log(t))}
\end{equation} 
as functions of time in a double logarithmic plot. 
As can be deduced from Fig.\ref{Fig1} (a,b), the initial diffusive regime where $\Delta(t) \propto t$ and the subsequent ballistic regime $\Delta(t) \propto t^2$ are clearly visible and reproduced by our short-time expansion. As expected, for large times there are 
increasing deviations between theory and simulation as  the theory is a short-time expansion, and this is especially noticeable for large values of $\epsilon^2 k^4$, as for example is shown in Fig.\ref{Fig1} (c,d).

\subsection{Underdamped active Langevin motion}
\label{Sub:UPot}

For macroscopic self-propelled particles or particles in a gaseous medium, inertial effects are getting relevant and 
overdamped active Brownian motion is generalized towards underdamped active Langevin motion \cite{scholz_inertial_2018,lowen_inertial_2020}.
The equations of motion for an inertial active particle in a random potential energy landscape are then generalized to
\begin{align}
\label{17}
m\ddot{\vec{r}}(t)+\gamma\dot{\vec{r}}(t)&= \gamma v_0\hat{u}(t)+\vec{F}(\vec{r}(t))+\vec{f}(t),\\
\label{18}
\gamma_R\dot{\phi}(t)&= f_R(t),
\end{align}
where $m$ is the particle mass. For simplicity, as in many previous studies for inertia \cite{enculescu_active_2011,takatori_inertial_2017,mokhtari_collective_2017,das_local_2019}, 
 we have neglected rotational inertia here 
which could be included by using a finite moment of inertia \cite{scholz_inertial_2018,lowen_inertial_2020}.

Now the initial condition  average $\ll \cdot \gg$ 
has to be performed not only over particle positions and orientations but also over 
the initial particle velocity $\dot{\vec{r}}(0)$.
 The resulting triple-averaged short-time expansion of the mean square displacement  is now:
\begin{align}
\label{19}
    \Delta(t)&=\sigma^2_v t^2+\frac{\gamma}{m}\left[\frac{4}{3}\frac{\gamma}{m}D-\sigma^2_v\right]t^3\nonumber\\
    &+\frac{1}{m^2}\left[\frac{7}{12}\gamma^2\sigma^2_v+\frac{1}{4}\widehat{F_i^2}+\frac{1}{4}\gamma^2v_0^2-\frac{\gamma^3}{m}D\right]t^4\nonumber\\    
    &+\mathcal{O}\left(t^5\right),
\end{align}
where $\sigma^2_v=\ll\dot{x}^2(0)+\dot{y}^2(0)\gg$ is the variance of the initial speed of the particle.
This result bears different dynamical scaling regimes. First of all, for short-times the MSD starts ballistically with $t^2$ due to the initial velocities. Of course this regime is absent if the particle is initially at rest when $\sigma^2_v=0$. 
Remarkably, for $\sigma^2_v \ll D\gamma/m$ the leading behavior is governed by the term $t^3$, {\it cubic\/} in time, as the prefactor is positive. Please note that for an initially thermalized particle with a Maxwellian velocity distribution, the prefactor is negative, implying the absence of this cubic regime. Finally, the presence of an external disordered force field now contributes to the  $t^4$ term as does the self-propulsion. 
This is plausible, as if on average a constant (external or internal self-propulsion) force is present, then the particle is constantly accelerated which leads to the $t^4$-scaling. 
Consequently, for  $\sigma^2_v \ll D\gamma/m \ll \overline{F_i^2}/\gamma^2+v_0^2$ there are {\it three} subsequent scaling regimes: from initially ballistic, over to the cubic regime and finally to the constant acceleration regime.\\
The typical crossover time between the $t^2$ and $t^3$ scalings and the one between $t^3$ and $t^4$ are referred to as $t^c_{2\rightarrow 3}$ and $t^c_{3\rightarrow 4}$. Their values are:
\begin{align}
\label{20}
t^c_{2\rightarrow 3}&=\frac{m}{\gamma}\frac{\sigma^2_v}{\frac{4}{3}\frac{\gamma}{m}D-\sigma^2_v},\\
\label{21}
t^c_{3\rightarrow 4}&=m\gamma\frac{\frac{4}{3}\frac{\gamma}{m}D-\sigma^2_v}{\frac{7}{12}\gamma^2\sigma^2_v+\frac{1}{4}\widehat{F_i^2}+\frac{1}{4}\gamma^2v_0^2-\frac{\gamma^3}{m}D},
\end{align}
where we assume that both prefactors of $t^3$ and $t^4$ in Eq.(\ref{19}) are positive.

\begin{figure}[!htbp]
 \begin{center}
 \includegraphics[width=8.7cm]{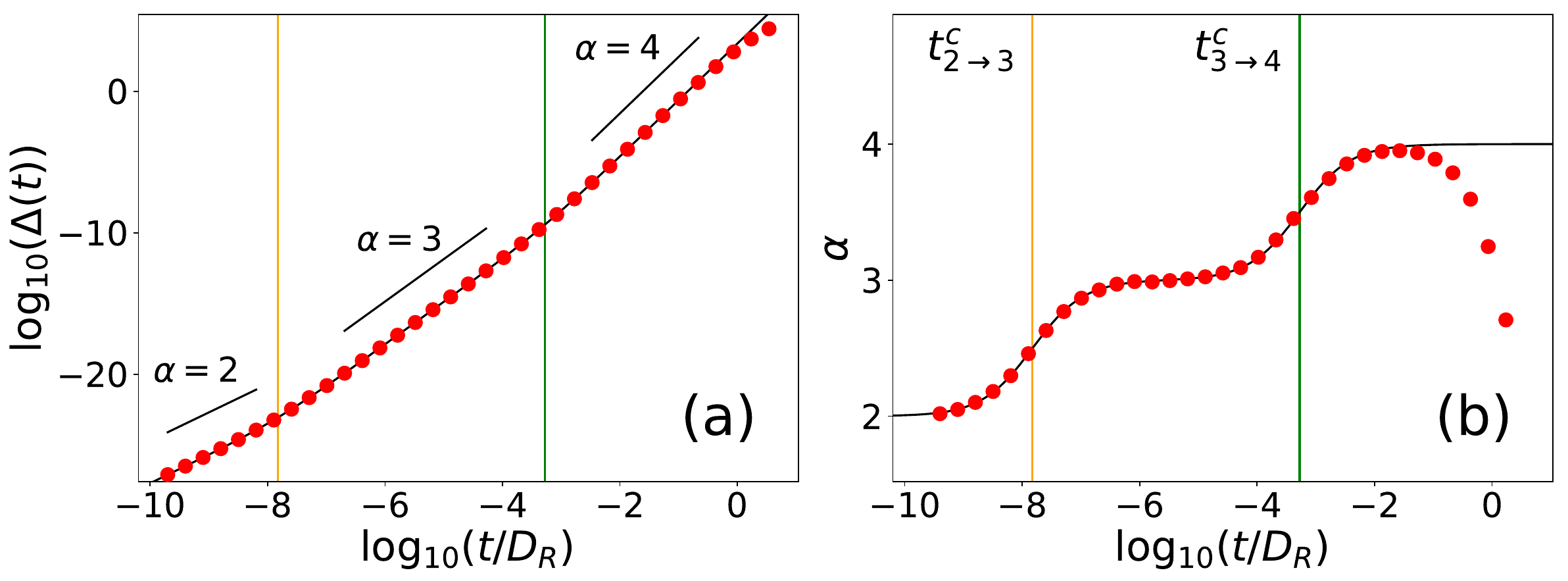}
 \caption{Mean square displacement (a) for an underdamped active particle in a random single mode potential, with scaling exponent $\alpha$ (b) and crossing times $t^c_{2\rightarrow 3}$, $t^c_{3\rightarrow 4}$. The parameters used are $v_0=100\sqrt{DD_R}$, $\epsilon=100D\gamma$, $L=100\sqrt{D/D_R}$ and $\sigma_v=0.0002\sqrt{DD_R}$, and the unit for mass is the mass of the particle $m$. The three different scalings $t^2$, $t^3$ and $t^4$ are in this case clearly distinguishable from each other.}
 \label{Fig2}
 \end{center}
\end{figure}

Using Langevin dynamics computer simulations, we have compared the theoretical short-time expansion with simulation data in Figure \ref{Fig2}. We used for the time evolution of the system a symmetrical stochastic splitting method that separates the stochastic and deterministic parts of the differential equations \cite{bussi_accurate_2007,sivak_using_2013}, with a typical time step of $\Delta t=10^{-10}/D_R$. As for the overdamped case, we used a single mode potential field and we averaged the MSD over more than $10^4$ configurations of the initial conditions and the potential.

A double-logarithmic plot indeed reveals three distinctive regimes where the MSD scales as $t^\alpha$ with $\alpha=2,3,4$ and there is good agreement between theory and simulation if the times are not too large. It is important to note that the cubic regime can only be seen for initially cool systems which are exposed to thermal fluctuations. 
These can be experimentally prepared for example
for granular hoppers \cite{scholz_inertial_2018} which are initially at rest and then brought into motion by instantaneously changing the vibration amplitude and frequency. Hence though the $t^3$ regime is not visible for a thermalized system it shows up for relaxational dynamics even for passive particles.

\section{Active particle in a disordered motility landscape}
\label{Sec:motility}

\subsection{No aligning torque, overdamped}
We now consider a self-propelling velocity that fluctuates \cite{zaburdaev_random_2008} as a function of the position of the particle. We denote hence the fluctuating part of the self-propelling velocity with $\delta v (\vec{r})$, while the constant part will still be named $v_0$, leading to a total propulsion velocity $(v_0+\delta v (\vec{r}))\hat{u}(\phi)$, or motility field. As in the case of the random potential, the random motility field is decomposed into two-dimensional Fourier modes, with Gaussian uncorrelated amplitudes: 

\begin{align}
\label{22}\delta v(\vec{r})=\sum_{i,j=0}^\infty\left(\zeta_{ij}^{(1)}\cos(k_ix+k_jy)+\zeta_{ij}^{(2)}\sin(k_ix+k_jy)\right),
\end{align}
where the $\zeta_{ij}^{(\alpha)}$ prefactors have the same statistical properties as the $\epsilon_{ij}^{(\alpha)}$ prefactors in (\ref{11}).\\
The main differences between the motility and potential fields are that the first one does not appear as a gradient in the equations of motion and that it is coupled to $\hat{u}(\phi)$.\\
In absence of an aligning torque and inertia the system fulfils the equations:

\begin{align}
\label{23}
\gamma\dot{\vec{r}}(t)&= \gamma(v_0+\delta v (\vec{r}))\hat{u}(\phi)+\vec{f}(t),\\
\label{24}
\gamma_R \dot{\phi}(t)&= f_R(t),
\end{align}
leading to the following short-time mean square displacement:

\begin{align}
\label{25}
    \Delta(t)&=4Dt+(v_0^2+\widehat{\delta v^2})t^2\nonumber\\
    &-\frac{1}{3}\left[2D\widehat{\delta v^{i2}}+D_R(v_0^2+\widehat{\delta v^2})\right]t^3\nonumber\\
    &+\frac{1}{24}\left[6D^2\widehat{\delta v^{ij2}}
     +8DD_R\widehat{\delta v^{i2}}+2D_R^2(v_0^2+\widehat{\delta v^2})\right. \nonumber\\
    &\left.+7\widehat{\delta v^2 \delta v^{i2}}+4\widehat{\delta v^3 \delta v^{ii}}-5v_0^2\widehat{\delta v^{i2}}\right]t^4+\mathcal{O}\left( t^5\right),
\end{align}
where we use the same notation as described for Eq.(\ref{13}): the symbol $\widehat{(\cdot)}$ indicates an average over disorder and initial conditions, while the superscripts of $\delta v$ indicate sums over derivatives.
 We also remark that the product $\widehat{\delta v^3 \delta v^{ii}}$ is negative, while all the others are positive.\\
From the results in Eq.(\ref{25}) we can extract similar considerations as those we discussed in \ref{Sub:OPot} for Eq.(\ref{13}). In the limit of a vanishing motility field $\delta v(\vec{r})=0$, the mean square displacement of an active particle with constant speed (see Eq.(\ref{14})) is recovered. For a finite total self-propulsion velocity the first correction to the linear MSD is a $t^2$ term which is always positive, leading to a ballistic regime. The typical crossover time related to this transition $t_{1\rightarrow 2}^c$ is now
\begin{equation}
\label{26}
t^c_{1\rightarrow 2}=\frac{4 D}{\widehat{\delta v^2}+v_0^2}.
\end{equation}

Similar to Eq.(\ref{13}), the space configuration of the field appears for the first time in the $\mathcal{O}\left(t^3\right)$ term of the equation as a negative term that does not constitute a regime. The $\mathcal{O}\left(t^4\right)$ prefactor is positive for a large motility field and a small $v_0$, but as the higher order terms always overshadow this, the particle never shows a pure accelerating behavior.

\begin{figure}[!htbp]
 \begin{center}
 \includegraphics[width=8.7cm]{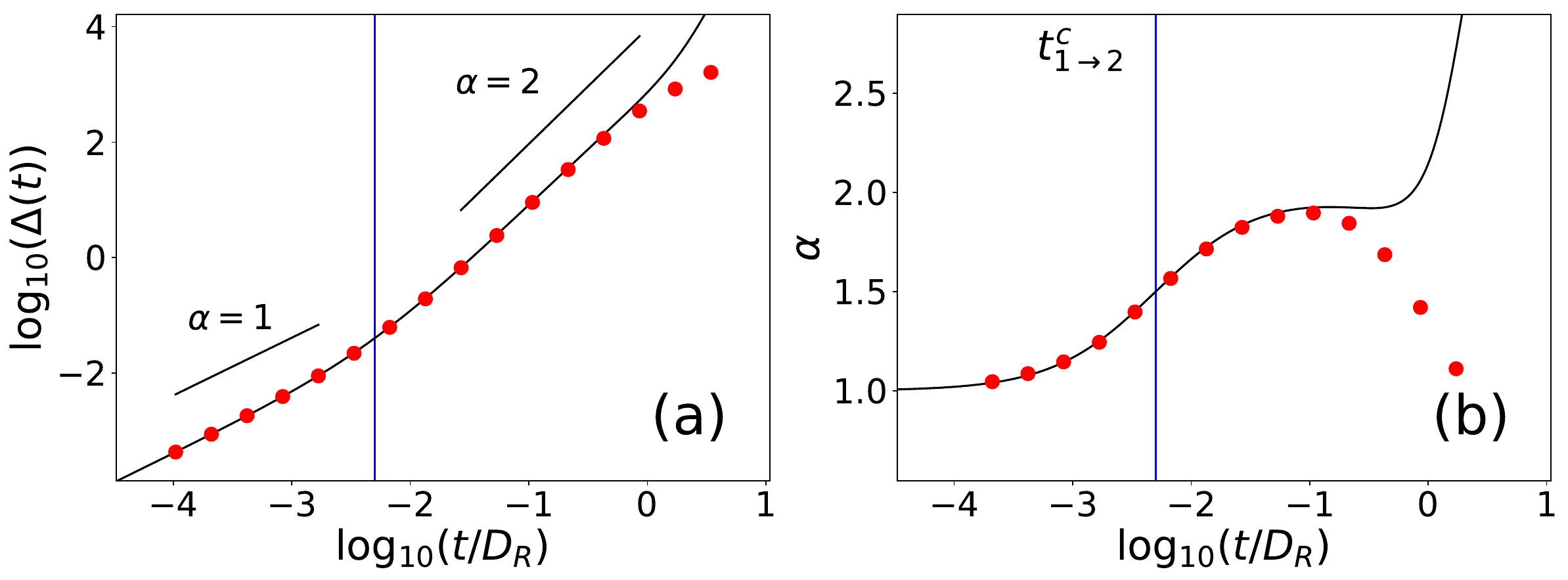}
 \caption{Mean square displacement (a), scaling exponent $\alpha$ (b) and crossing time $t^c_{1\rightarrow 2}$ for an underdamped active particle in a random single mode motility field. The parameters $v_0=20\sqrt{DD_R}$ and $\zeta=20\sqrt{DD_R}$, $L=100\sqrt{D/D_R}$ feature the initial diffusive behavior and the ballistic behavior.}
\label{Fig3}
\end{center}
\end{figure}

All these results have been confirmed by simulations similar to those described in \ref{Sub:OPot}. In Figure \ref{Fig3} we can see an example of such a simulation, where the plots of the MSD and its scaling exponent $\alpha$ behave in accord to our theory for short-times, with first a diffusive regime and then a ballistic one.

\subsection{No aligning torque, underdamped}
The underdamped equations of motion for a massive particle subject to a random motility field and no aligning torque are:

\begin{align}
\label{27}
m\ddot{\vec{r}}(t)+\gamma\dot{\vec{r}}(t)&= \gamma(v_0+\delta v (\vec{r}))\hat{u}(\phi)+\vec{f}(t),\\
\label{28}
\gamma_R\dot{\phi}(t)&=f_R(t),
\end{align}
we ignore the effects of angular inertia, for the same reason explained in \ref{Sub:UPot}.

The resulting MSD, averaged over disorder, initial conditions and thermal noise is:
\begin{align}
\label{29}
    \Delta(t)&=\sigma^2_v t^2+\frac{\gamma}{m}\left[\frac{4}{3}\frac{\gamma}{m}D-\sigma^2_v\right]t^3\nonumber\\
    &+\frac{\gamma^2}{m^2}\left[\frac{7}{12}\sigma^2_v+\frac{1}{4}(v_0^2+\widehat{\delta v^2})-\frac{\gamma}{m}D\right]t^4\nonumber\\    
    &+\mathcal{O}\left(t^5\right).
\end{align}
The three consecutive scaling regimes that characterized Eq.(\ref{19}): $t^2$, $t^3$ and $t^4$, can be also found in Eq.(\ref{29}) by requiring now $\sigma^2_v \ll D\gamma/m \ll \widehat{\delta v^2}+v_0^2$. The crossing time $t_{3\rightarrow 4}$ changes accordingly, while $t_{2\rightarrow 3}$ remains the same that we calculated in the potential case (see Eq.(\ref{20})):

\begin{align}
\label{30}
t^c_{2\rightarrow 3}&=\frac{m}{\gamma}\frac{\sigma^2_v}{\frac{4}{3}\frac{\gamma}{m}D-\sigma^2_v},\\
\label{31}
t^c_{3\rightarrow 4}&=\frac{m}{\gamma}\frac{\frac{4}{3}\frac{\gamma}{m}D-\sigma^2_v}{\frac{7}{12}\sigma^2_v+\frac{1}{4}(v_0^2+\widehat{\delta v^2})-\frac{\gamma}{m}D},
\end{align}
where we assume that both the prefactors of $t^3$ and $t^4$  in Eq.(\ref{29}) are positive.

\begin{figure}[!htbp]
 \begin{center}
 \includegraphics[width=8.7cm]{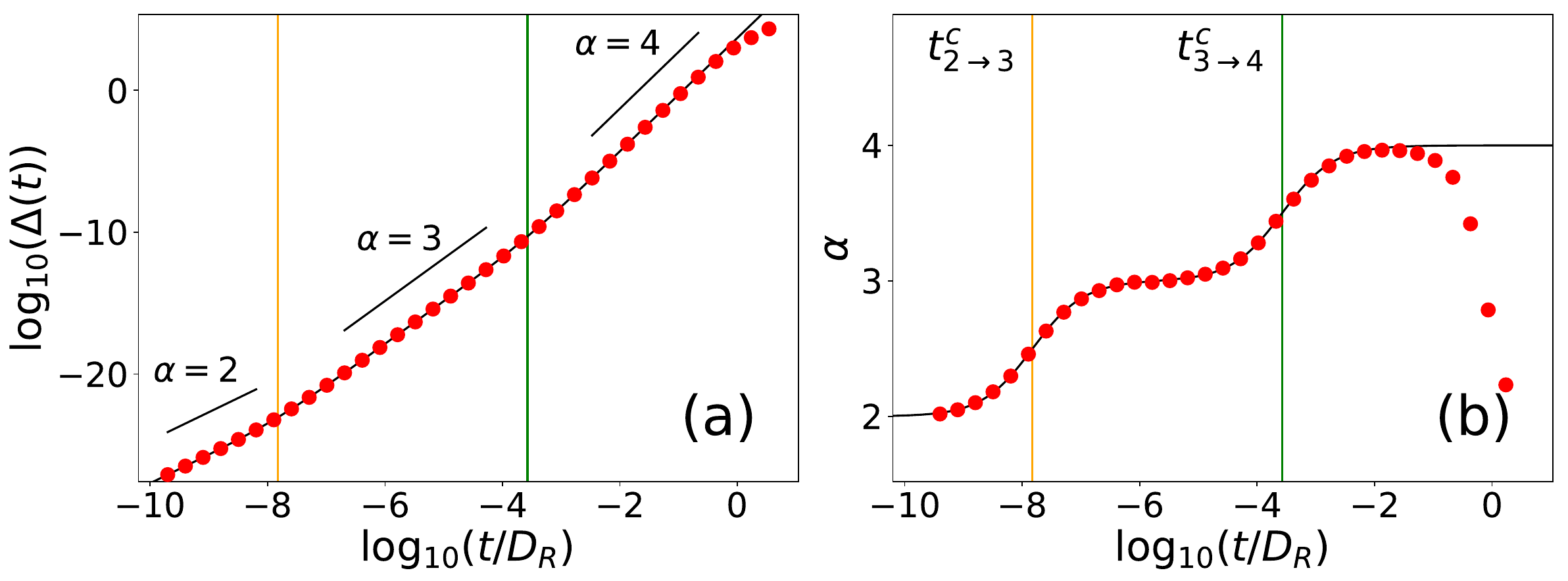}
 \caption{Mean square displacement (a) for an underdamped active particle in a random single mode motility field, with scaling exponent $\alpha$ (b) and crossing times $t^c_{2\rightarrow 3}$, $t^c_{3\rightarrow 4}$. The parameters used are $v_0=100\sqrt{DD_R}$, $\zeta=100\sqrt{DD_R}$, $L=100\sqrt{D/D_R}$ and $\sigma_v=0.0002\sqrt{DD_R}$, and the unit of mass is the mass of the particle $m$. The three different scalings $t^2$, $t^3$ and $t^4$ are clearly distinguishable.}
 \label{Fig4}
 \end{center}
\end{figure}

These results were compared to the numerical MSD calculated with the help of Langevin dynamics simulations. In Figure \ref{Fig4} we present the typical results that can be obtained when the limit $\sigma^2_v \ll D\gamma/m \ll \widehat{\delta v^2}+v_0^2$ applies, and hence three different regimes appear.

\subsection{Aligning torque}
In this subsection we discuss the special case of the presence of an aligning torque $\tau(\vec{r},\phi)$ that redirects the self-propulsion of the particle towards either the maxima or the minima of the motility field. An aligning torque is important for colloidal realizations of active systems \cite{lozano_phototaxis_2016,jahanshahi_realization_2020,jahanshahi_colloidal_2019,geiseler_self-polarizing_2017}. Since one common way of realizing a motility field is by the use of light fields, we refer to the self-propulsion towards the maxima of the field as \emph{positive phototaxis} and the one towards the minima as \emph{negative phototaxis}.

Here, we only focus on the underdamped case, characterized by the following equations:

\begin{align}
\label{32}
\gamma \dot{\vec{r}}(t)&= \gamma (v_0+\delta v (\vec{r}))\hat{u}(\phi)+\vec{f}(t),\\
\label{33}
\gamma_R\dot{\phi}(t)&=\gamma_R \tau(\vec{r},\phi)+f_R(t),
\end{align}
where $\tau(\vec{r},\phi)\equiv q(v_0+\delta v(\vec{r}))\left(\vec{\nabla}\delta v(\vec{r})\times\vec{u}(\phi)\right)\cdot \vec{e}_z$. The sign of the prefactor $q$  determines whether the phototaxis is positive ($q<0$) or negative ($q>0$).\\
The averaged MSD up to $\mathcal{O}\left( t^4\right)$ is:
\begin{align}
\label{34}
    \Delta(t)&=4Dt+(v_0^2+\widehat{\delta v^2})t^2\nonumber\\
    &-\frac{1}{3}\left[2D(1+qv_0)\widehat{\delta v^{i2}}+D_R(v_0^2+\widehat{\delta v^2})\right]t^3\nonumber\\
    &+\mathcal{O}\left( t^4\right).
\end{align}
In the special case of no translational diffusion ($D=0$) the next order of the MSD is:

\begin{align}
\label{35}
    \Delta(t)&=\dots+\frac{1}{24}\left[ 2D_R^2(v_0^2+\widehat{\delta v^2})+7\widehat{\delta v^2 \delta v^{i2}}-5v_0^2\widehat{\delta v^{i2}}\right. \nonumber\\
    &+4\widehat{\delta v^3 \delta v^{ii}}-4q(v_0^3\widehat{\delta v^{i2}}+3v_0\widehat{\delta v^2\delta v^{i2}})\nonumber\\
    &\left.+3q^2(v_0^4\widehat{\delta v^{i2}}+6v_0^2\widehat{\delta v^2\delta v^{i2}}\widehat{\delta v^4\delta v^{i2}})\right]t^4\nonumber\\
    &+\mathcal{O}\left( t^5\right).
\end{align}
Analyzing Equations (\ref{34}) and (\ref{35}) we first notice that in the limit of $q=0$ we recover the previous case with no aligning torque. When $q$ is non-zero, it appears for the first time as prefactor of $t^3$ if $D>0$ and as prefactor of $t^4$ otherwise. What is peculiar about $q$ is that for different experimental setups its sign can change, and when it is negative, all the prefactors where it appears become positive. One can intuitively understand the reason for this by considering that a positive phototaxis means that the particle redirects itself towards the motility field maxima, and hence will show a MSD which is larger than in the negative phototaxis case. Even when $q$ is negative and large though, this does not constitute a regime of either order $t^3$ or $t^4$, as the higher order terms in time feature higher powers of $q$ that overshadow the lower orders.

\section{Conclusions and outlook}
\label{Sec:conclusions}
In conclusion we have systematically computed the quenched  disorder average of the  mean-square-displacement
for an active particle in a random potential or motility landscape. The amplitude of the ballistic regime is affected by the strength of disorder but spatial derivatives in the landscapes only contribute to the next cubic term in time.
For an inertial particle two new superballistic scaling regimes are found where the MSD scales as $t^3$ or as $t^4$.

Our method can be applied to other more complex situations \cite{woillez_active_2020}. First, the generalization to an anisotropic potential is straightforward, even though tedious.
Second, the landscapes can be time-dependent
as for real speckle patterns \cite{paoluzzi_run-and-tumble_2014}, moving activity waves \cite{geiseler_self-polarizing_2017,merlitz_linear_2018} and propagating ratchets
\cite{lozano_propagating_2019,zampetaki_taming_2019,koumakis_dynamic_2019}
The same analysis can be performed for time-dependent disorder. 
Moreover, the same analysis can in principle be done for other models of active particles, 
including the simpler  active Ornstein Uhlenbeck particle  \cite{martin_statistical_2020} or more sophisticated pusher or puller descriptions for the self-propagation. A refreshing or resetting of the landscapes can be considered 
as well \cite{mano_optimal_2017,scacchi_mean_2018}. Finally the model can be extended to a viscoelastic solvent 
\cite{gomez-solano_dynamics_2016,berner_oscillating_2018,qi_enhanced_2020,theeyancheri_translational_2020}
with a random viscoelasticity where memory effects become important.

\section{Acknowledgements}
We thank S. U. Egelhaaf
and C. Zunke
for helpful discussions. The work of DB was supported within the EU  MSCA-ITN ActiveMatter,
(proposal No.\ 812780). HL acknowledges funds from the German Research Foundation (DFG) within SPP 2265 within project LO 418/25-1.

\appendix
\section{Example of mean square displacement calculation}
\label{Appendix}
In this appendix we present an example for how we calculated the analytical results in this paper. Specifically, we will show the procedure used for the case of an overdamped particle in a random potential (see Eq.(\ref{13})).

\subsection{Model system}

The equation of motion for the time dependent position $x(t)$ of the particle is given by Equations (\ref{1}) and (\ref{2}). Taylor-expanding $\vec{F}(\vec{r}(t))$ around the starting position $\vec{r}(0)\equiv\vec{r}_0$ yields
\begin{align}
\label{A1}
\vec{F}(\vec{r}(t)) &= \sum_{n_x=0}^\infty \sum_{n_y = 0}^\infty  \frac{(x(t)-x_0)^{n_x} (y(t)-y_0)^{n_y}}{n_x! n_y!} \nonumber\\
&\times\left(\frac{\partial^{n_x + n_y}\vec{F}}{\partial x^{n_x}\partial y^{n_y}}\right)(\vec{r}_0).
\end{align}
We truncate this expression in the following way:
\begin{equation}
\label{A2}
\vec{F}(\vec{r}(t))\simeq
\begin{pmatrix}
F_x(\vec{r}_0)+F_x^x(\vec{r}_0)(x(t)-x_0)\\
F_y(\vec{r}_0)+F_y^y(\vec{r}_0)(y(t)-y_0)
\end{pmatrix},
\end{equation}
where a subscript in $F$ denotes a component of the force and a superscript indicates a partial derivative.\\
In this way we approximate our system to an active particle subject to two Brownian oscillators in the $x$ and $y$ directions independent of each other. The additional force terms of higher order will be treated in perturbation theory. The goal is to calculate the mean square displacement $\Delta(t):=\ll\overline{\left< (\vec{r}(t) - \vec{r}_0)^2 \right>}\gg$ for short-times up to forth order in time but for arbitrary strength of the random forces.

\subsection{Active Brownian oscillator}

We will focus on the equation in the $x$ component, as the one in $y$ can be treated in an analogous way. First we consider the formal solution of the active Brownian oscillator
\begin{align}
\label{A3}
 \gamma \dot{x}_{B} &=   f_x(t)+\gamma v_0\cos(\phi(t))\nonumber\\
 &+F_x(\vec{r}_0)+ F_x^x(\vec{r}_0)(x_{B}(t)-x_0),
\end{align}
which is
\begin{align}
\label{A4}
x_{B}(t) &= x_0\nonumber +\frac{F_x(x_0)}{F_x'(x_0)}\left(e^{\frac{1}{\gamma} F_x'(x_0)t}-1\right) \\&\ +\frac{1}{\gamma}\int_0^te^{\frac{1}{\gamma} F_x'(x_0)(t-t')}f_x(t')dt'\nonumber \\&\ +v_0\int_0^te^{\frac{1}{\gamma} F_x'(x_0)(t-t')}\cos(\phi(t'))dt'
\nonumber \\&\ =:x_0+x_a(t)+x_b(t)+x_c(t),
\end{align}
where 
\begin{equation}
\label{A5}
\phi(t)=\frac{1}{\gamma_R}\int_0^t f_R(t') dt.
\end{equation}
The mean square displacement in the $x$ direction is
\begin{align}
\label{A6}
\Delta_{xB}(t)&= \ll\overline{\left<(x_a(t)+x_b(t)+x_c(t))^2\right>}\gg\nonumber \\
&=2Dt+\left(\frac{\widehat{F_x^2}}{\gamma^2}+\frac{v_0^2}{2}\right)t^2+\frac{1}{6}\left(
%
8
D\frac{\widehat{F_x^{x2}}}{\gamma^2}-D_Rv_0^2\right) t^3\nonumber \\&\ +\frac{1}{24}\left(14
\frac{\widehat{F_x^2F_x^{x2}}}{\gamma^4}+7\frac{\widehat{F_x^{x2}}}{\gamma^2}v_0^2+D_R^2v_0^2\right) t^4+{\mathcal O}(t^5).
\end{align}
Note that we omitted all averages over odd powers of the force or its derivatives, as they are all accompanied by odd functions in space that go to zero when averaging over the initial conditions.

\subsection{Perturbation approach}

Now we will treat the time perturbation, considering terms up to fourth order in time. In order to do this, we will have to consider all the terms in Eq.(\ref{A3}) for which $n_x+n_y\leq 4$. 

We want to determine the solution
\begin{align}
\label{A7}
x(t)=x_B(t) +h_x^{(1)}(t)
\end{align}
that fulfils the following differential equation:
\begin{align}
\label{A8}
&\gamma \dot{x}_B(t)+\gamma \dot{h}_1(t)=f_x(t)+\gamma v_0\cos(\phi(t))\nonumber \\&\ +\sum_{n_x=0}^4 \sum_{n_y = 0}^4  \frac{(x_B(t)+h_x^{(1)}-x_0)^{n_x} (y_B(t)+h_y^{(1)}-y_0)^{n_y}}{n_x! n_y!}\nonumber\\
&\times \left(\frac{\partial^{n_x + n_y}F_x}{\partial x^{n_x}\partial y^{n_y}}\right)(\vec{r}_0).
\end{align}
If we consider a small perturbation $h_x^{(1)}(t)$, we obtain:
\begin{align}
\label{A9}
\gamma h^{(1)}_x(t)&\simeq\int_0^t\left[F_x^y(\vec{r}_0)(y_B(t')-y_0)\right.\nonumber\\
&+\frac{F_x^{xx}(\vec{r}_0)}{2}(x_B(t')-x_0)^2\nonumber\\
&+F_x^{xy}(\vec{r}_0)(x_B(t')-x_0)(y_B(t')-y_0)\nonumber\\
&+\frac{F_x^{yy}(\vec{r}_0)}{2}(y_B(t')-y_0)^2\nonumber\\
&+\left.\frac{F_x^{xxx}(\vec{r}_0)}{6}(x_B(t')-x_0)^3+\dots\right]dt'
\end{align}
where we first used the differential equation of the unperturbed Brownian oscillator and then assumed that $h_x^{(1)}(t)$ is small. The fifth order derivatives of the force have been omitted because they would not lead to any terms of forth or smaller order in $t$.	\\
Similarly we calculate the second order perturbation $h_x^{(2)}(t)$, while higher order perturbations are not necessary.

The mean square displacement within the first and second order perturbation theory is
\begin{align}
\label{A10}
\Delta_x(t)&=\ll \overline{\left<\left(x_{a}(t)+x_{b}(t)+h_x^{(1)}+h_x^{(2)}(t)\right)^2\right>}\gg,
\end{align}
and the only thing left is to explicitly calculate this expression and sum it to the respective one for the $y$ direction.

\subsection{Simplification of averages}
Given the potential described in Eq.(\ref{10}), one is able to simplify the various expressions for the averages of the forces and their derivatives.
For example we have: 
\begin{align}
\label{A11}
\widehat{F_x^2} &= \frac{1}{2}\sum_{i,j,\alpha} \overline{\epsilon_{ij}^{(\alpha)2}}k_i^2, \\
\label{A12}
\widehat{F_x^{x2}} &= \frac{1}{2}\sum_{i,j,\alpha} \overline{\epsilon_{ij}^{(\alpha)2}}k_i^4 ,\\
\label{A13}
\widehat{F_xF_x^{xx}} &= -\frac{1}{2}\sum_{i,j,\alpha} \overline{\epsilon_{ij}^{(\alpha)2}}k_i^4 =-\widehat{F_x^{x2}},\\
\label{A14}
\widehat{F_x^2F_x^{y2}} &= \frac{3}{8} \sum_{i,j,\alpha} {\overline{\epsilon_{ij}^{(\alpha)4}}}k_i^4k_j^2 + \frac{1}{4} \sum_{\substack{i\neq m \land j\neq n \\ \alpha,\beta}} \overline{\epsilon_{ij}^{(\alpha)2}}\,\overline{\epsilon_{mn}^{(\beta)2}} k_i^2 k_m^2k_n^2,\\
\label{A15}
\widehat{F_x^2F_y{F_x^{xy}}} &=-\widehat{F_x^2F_x^{y2}}, \\
\text{etc}\dots\nonumber
\end{align}
Using these relations we can write the whole expression for the MSD using only terms that we know to be positive. One has to be careful though, especially with the products containing four terms, as for example in Eq.(\ref{A16}). These kind of products contain both a common mode average and a cross mode one (for example respectively the first and second sum in Eq.(\ref{A16})). It can happen that two different products contain the same (or opposite) common mode average but a different cross mode one. For example:
\begin{align}
\label{A16}
\widehat{F_x^2F_x^{x2}} &= \frac{3}{8} \sum_{i,j,\alpha} {\overline{\epsilon_{ij}^{(\alpha)4}}}k_i^6 + \frac{1}{4} \sum_{\substack{i\neq m \land j\neq n \\ \alpha,\beta}} \overline{\epsilon_{ij}^{(\alpha)2}}\,\overline{\epsilon_{mn}^{(\beta)2}} k_i^4 k_m^2,\\
\label{A17}
\widehat{F_x^3F_x^{xx}} &=- \frac{3}{8} \sum_{i,j,\alpha} {\overline{\epsilon_{ij}^{(\alpha)4}}}k_i^6 - \frac{3}{4} \sum_{\substack{i\neq m \land j\neq n \\ \alpha,\beta}} \overline{\epsilon_{ij}^{(\alpha)2}}\,\overline{\epsilon_{mn}^{(\beta)2}} k_i^4 k_m^2.
\end{align}
In this case the absolute value of the cross mode of (\ref{A19}) is three times larger than that of (\ref{A18}), while the common mode is the same. In other cases, these cross modes can even disappear:
\begin{align}
\label{A18}
\widehat{F_x^2F_x^{y2}} &= \frac{3}{8} \sum_{i,j,\alpha} {\overline{\epsilon_{ij}^{(\alpha)4}}}k_i^4k_j^2 + \frac{1}{4} \sum_{\substack{i\neq m \land j\neq n \\ \alpha,\beta}} \overline{\epsilon_{ij}^{(\alpha)2}}\,\overline{\epsilon_{mn}^{(\beta)2}} k_i^2 k_m^2k_n^2,\\
\label{A19}
\widehat{F_xF_yF_x^xF_x^y} &= \frac{3}{8} \sum_{i,j,\alpha} {\overline{\epsilon_{ij}^{(\alpha)4}}}k_i^4k_j^2.
\end{align}
In the special case of a single mode potential the following expression of Eq.(\ref{13}):
\begin{align}
\label{A20}
&14\widehat{F_i^2F_i^{i2}}+8\widehat{F_i^3F_i^{ii}}+14\widehat{F_xF_yF_x^{y}F_i^{i}}\nonumber\\
& +14\widehat{F_yF_xF_y^{x}F_i^{i}} -5\widehat{F_i^2F_x^{y2}}-5\widehat{F_i^2F_y^{x2}}
\end{align}
simplifies to $6\widehat{F_i^2F_j^{k2}}$.

\end{document}